\documentclass[12pt, epsf]{article} 
\input epsf.tex 
\def\be{\begin{equation}} \def\ee{\end{equation}}
\def\bi{\begin{itemize}} \def\ei{\end{itemize}}
\def\bea{\begin{eqnarray}} \def\eea{\end{eqnarray}} \def\ba{\begin{array}}
\def\ea{\end{array}} \def\ben{\begin{enumerate}} \def\een{\end{enumerate}}

\newcommand{\eqn}[1]{(\ref{#1})}

\newcommand{\prl}[3]{Phys. Rev. Lett. {\bf#1} ({#2}) {#3}}

\newcommand{\hepth}[1]{{\tt arXiv:{#1}[hep-th]}}

\def\br{\nonumber\\}
 
\def\g{\gamma}

\begin{document}
{}~
\hfill \vbox{
\hbox{arXiv:1507.nnnn} 
\hbox{\today}}\break

\vskip 3.5cm
\centerline{\Large \bf
Perturbative entanglement thermodynamics 
}
\centerline{\Large \bf
for AdS spacetime: Renormalization
}

\vskip 1cm

\vspace*{1cm}

\centerline{\sc  
Rohit Mishra and Harvendra Singh
}

\vspace*{.5cm}
\centerline{ \it  Theory Division, Saha Institute of Nuclear Physics} 
\centerline{ \it  1/AF Bidhannagar, Kolkata 700064, India}
\vspace*{.25cm}

\vspace*{.5cm}

\vskip.5cm
%\centerline{E-mail: h.singh [AT] saha.ac.in }

\vskip1cm
%DRAFT : \today \\

\centerline{\bf Abstract} \bigskip
We study the effect of charged excitations   in the AdS spacetime
on the first law of  entanglement thermodynamics.
It is found that  `boosted' AdS black holes
give rise to a more  general form of first law which  includes 
chemical potential and  charge density. To obtain this result
we have to resort to a second order perturbative 
calculation of entanglement entropy for small size subsystems. 
At  first order the form  of entanglement  law 
remains unchanged even in the presence of charged excitations.
But  the thermodynamic quantities  
have to be appropriately `renormalized' at the second 
order due to the corrections. We work in the perturbative 
regime where $T_{thermal}\ll T_E$.

\vfill 
\eject

\baselineskip=16.2pt

%%%%%%%%%%%%%%%%%%%%%%%%%%%%%%%%%%%%%%%%%%
%%%%%%%%%%%%%%%%%%%%%%%%%%%%%%%%%%%%%%%%%%%%%%

\section{Introduction}

The AdS/CFT correspondence \cite{malda} has been a very
successful idea in string theory. It relates conformal field theries
living on the boundary of  anti de Sitter (AdS)
spacetime with the supergravity theory in the bulk. 
The idea of entanglement entropy
has also been a focus of recent study 
in string theory \cite{RT}.
It has led to some understanding of entanglement entropy in strongly 
coupled quantum mechanical systems, 
particularly  theories which exhibit scaling property near 
the critical points \cite{ogawa}. 
A significant observation has been 
that the small excitations of the subsystems in
the boundary theories follow entanglement thermodynamic  laws similar to the
black hole thermodynamics at finite temperature \cite{JT,alisha,Pang:2013lpa}.\footnote{  
See  also \cite{Ben-Ami:2014gsa} for work on multiple strips.} 
These calculations have become possible now because
 entanglement entropy can be estimated using the gauge/gravity holography 
\cite{RT}, that is
 by evaluating the geometrical area of  extremal sufaces embedded 
inside the bulk (AdS) geometry. It has been proposed recently 
in \cite{JT} that the entanglement entropy ($S_E$) and 
the energy of excitations (${\cal E}$) in pure AdS background give rise to
a relation
$$\bigtriangleup { \cal E}
= T_E \bigtriangleup S_E  +\cdots 
$$
which has been described as the first law of entanglement 
thermodynamics \cite{JT}.

In this paper we  study the effects of IR deformations (excitations)
 in asymptotically AdS spacetime which  carry gauge charges 
and look for modifications in the  entanglement first law. 
We do find that the `boosted' AdS black holes
give rise to a more general form of first law which includes the
chemical potential and  charge density. To obtain
 this result we have to resort to a second order perturbative 
calculation of the entanglement entropy. We  
find that various first order 
thermodynamic quantities, such as entropy, energy, temperature, etc 
get corrected  and these have to be suitably
redefined or `renormalized' at the second 
order. 
 The effects of higher order corrections appears similar to the renormalization 
procedure in quantum field theories. For example the strip width (subsystem size)
and entanglement temperature ($T_E$) 
have to be redefined to include corrections
so that a first law like relation holds good. Since we resort 
to perturbative calculations, 
we work in the regime where the ratio ${l \over z_0}$, of 
the strip width ($l$) to the horizon radius 
($z_0$), is kept very small. This 
hierarchy of scales can also be thought of in  terms of respective
temperatures as a limit 
$${T_{thermal}\over T_E} \ll {a_1\over 2b_0\gamma} \ .$$ 
We  mention that the  corrections to the entanglement entropy 
evaluated order by order  in  (dimensionless)
quantity  ${T_{thermal}\over T_E}$
should not be confused with (stringy) quantum corrections 
to the entanglement entropy
\cite{Faulkner:2013ana}. 

The paper is   organised as follows. In the section-2 we mainly work out 
the first law of entanglement thermodynamics
for AdS black holes in presence of boost. We get the familiar form of the law
which has the same temperature as in the unboosted case,
eventhough the charged excitations are present, while
the energy of excitations  increases compared to the unboosted case.
We  work out the entanglement entropy up to second order in  section-3. 
It is found that the form of first law changes 
under these corrections. The chemical potential explicitly appears in the 
first law at the second order. The temperature and other  
thermodynamic quantities need to be renormalized. 
In fact the entanglement temperature slightly
decreases with corrections. 
The conclusions are given in the section-4.

\section{Entanglement  from boosted AdS black holes}

The boosted $AdS_{d+1}$ black holes backgrounds  are
given by
\bea\label{bst1}
&&ds^2={L^2\over z^2}\left( -{f dt^2\over K}+K (dy-\omega)^2
+dx_1^2+\cdots+dx_{d-2}^2+{dz^2 \over f}\right) 
\eea
with
\be
 f=1-{z^d\over z_0^d}, ~~~~~K=1+\beta^2\gamma^2{z^d\over z_0^d} 
\ee
 $z_0$ is the horizon and  $0\le \beta\le 1$ is  boost parameter, while
 $\gamma={1\over\sqrt{1-\beta^2}}$. The boost is taken along $y$ direction.
The Kaluza-Klein form 
\be
\omega={\beta^{-1}}(1-{1\over K}) dt
\ee
and  $L$ is the radius of curvature of  AdS spacetime, which is very large.\footnote{
For example, in the $AdS_5\times S^5$ 
near-horizon geometry of $n$ coincident D3-branes, 
we shall have $L^4\equiv 2\pi g_{YM}^2 n$ and
the 't Hooft coupling constant $g_{YM}^2 n \gg 1$.}   

We study the entanglement entropy of a subsystem on the 
boundary of  boosted $AdS_{d+1}$ backgrounds in \eqn{bst1}.
We  embed  a $(d-1)$-dimensional strip-like  spatial surface,
in the bulk asymptotic geometry. 
The boundaries of the
extremal bulk surface  coincide with the two ends 
of the interval $-l/2 \le x^1 \le l/2$.   
The regulated  size of the
rest of the coordinates is taken large
$0\le x^i\le l_i$, with $l_i\gg l$. 
We shall always have coordinate $y$  being compact, so that
$0\le y\le 2\pi r_y$. As per the Ryu-Takayanagi prescription \cite{RT}
the entanglement entropy   is given 
in terms of the geometrical area of the extremal surface 
\bea\label{schkl1saa}
 S_E  &\equiv& {[A]_{Strip}\over 4G_{d+1}} \br
&=&
 { V_{d-2}  L^{d-1} \over 
2G_{d+1}}
\int^{z_\ast}_{\epsilon}{dz\over z^{d-1}} \sqrt{K} 
\sqrt{{1\over {f}}  +({\partial_z x^1})^2}
\eea  
where  
$G_{d+1}$ is $(d+1)$-dimensional Newton's constant 
and $V_{d-2} \equiv (2\pi r_y) l_2 l_3 \cdots l_{d-2}$ is 
the  spatial volume of the  boundary. We will be mainly working for $d>2$.
In our notation $\epsilon\sim 0$ denotes the cut-off scale near the boundary
to regularize the UV divergences, 
and $z_\ast$ is the turning point
of extremal surface  inside the bulk geometry. In the above $K(z),~f(z)$
are  known functions, so we only need to solve for $x^1$. 
From \eqn{schkl1saa} it follows that  
a minimal surface will have to satisfy 
\bea\label{kl3}
{dx^1\over dz}\equiv   {({z\over z_c})^{d-1}}{1\over
 \sqrt{f}\sqrt{K-({z\over z_c})^{2d-2}}} 
\eea 
The constant $z_c$  is given by
  the turning point   relation
\be
{K_\ast}- ({z_\ast \over z_c})^{2d-2}=0 
\ee
where  $K_\ast=K(z)|_{z=z_\ast}$.
The identification of the boundary $x^1(0)=l/2$ leads to the integral
relation 
\be\label{klop}
{l \over 2}=  \int_0^{z_\ast} dz
({z\over z_\ast})^{d-1}{1 \over \sqrt{f} 
 \sqrt{{K\over K_\ast}- ({z\over z_\ast})^{ 2d-2}}} 
\ee
which relates $l$ with $z_\ast$, the turning point.   
The turning-point takes the mid-point value $x^1(z_\ast)=0$ on the boundary. 
The  expression of the entanglement entropy
for these boosted  AdS black hole  solutions becomes
\be\label{kl1kv}
S_{E}=
{  V_{d-2} L^{d-1} \over 2 G_{d+1}}
\int^{z_\ast}_{\epsilon}{dz \over  z^{d-1}}{K 
\over\sqrt{f}\sqrt{K- K_\ast({z\over z_\ast})^{2d-2}}} 
\ee  
The expression \eqn{kl1kv}
mathematically provides the  entanglement entropy 
for a strip-like subsystem on the  boundary. For pure AdS spacetime 
($z_0\to\infty,~f=1=K$)
these integrals can be evaluated  exactly \cite{RT}, but in the
 presence of black hole 
it is difficult to find analytical answers from  the integral \eqn{kl1kv}, 
although numerical estimates can always be made.

\subsection{Thin strip approximation}
In the cases where the  strip subsystem is a small part of a big system,
so that the turning 
point lies in the proximity of
 asymptotic boundary region only ($z_\ast \ll z_0$), 
one can evaluate the entanglement entropy integral
\eqn{kl1kv} by expanding it around its pure AdS value (treating pure AdS
as a ground state). We shall  take boost to be finite but small
$\beta\gamma \sim 1$. Under these approximations
the strip width equation \eqn{klop} can be expanded perturbatively as
\bea \label{dfg1}
{l}
&=& 2 \int_0^{z_\ast} dz
({z\over z_\ast})^{d-1}{1 \over \sqrt{f} 
 \sqrt{{K\over K_\ast}- ({z\over z_\ast})^{ 2d-2}}}  \br
&& 
= 2z_\ast \int_0^{1} d\xi
\xi^{d-1}{1 \over  
 \sqrt{R}}[ 1+ {1\over 2}{z_\ast^d\over z_0^d}\xi^d + 
{\beta^2\gamma^2 z_\ast^d \over 2z_0^d} {1-\xi^d\over R} +\cdots ]  \br
&&\equiv
2z_\ast\left( b_0 + 
{z_\ast^{d}\over 2 z_0^d}( b_1 + \beta^2\g^2 I_l)\right)  
+\cdots
\eea
where we have introduced $\xi={z\over z_\ast}$, $R\equiv 1- \xi^{ 2d-2} $ 
and the dots indicate terms of higher order in 
$({z_\ast\over  z_0})^{d}$. The coefficients $b_0, b_1$, and $ I_l$  are 
precise integral Beta functions  multiplying at  various orders.
These coefficients are provided in the appendix.
Note   $b_0$ and $b_1$ are positive definite quantities.
Keeping only up to  first order in $(z_\ast^d/z_0^d)$ 
the equation \eqn{dfg1} can be inverted to obtain
\bea\label{dfg5}
 z_{\ast}&=&{l/2  \over  b_0 + 
{z_\ast^{d}\over 2 z_0^d}( b_1 + \beta^2\g^2  I_l)} 
 \simeq {\bar z_\ast \over  
1 + {\bar z_\ast^{d}\over 2 z_0^d}( {b_1\over b_0} 
   + {\beta^2\g^2 \over  b_0}I_l)} 
\eea
where  
$\bar z_\ast\equiv {l  \over 2 b_0}$  being the turning point value 
in  pure AdS case having the same strip width  $l$. \footnote{ 
Using a function 
$g(z)=1+{z^d\over z_0^d}\simeq {1\over f}$ we may also express
$
z_\ast 
 \simeq 
{\bar z_{\ast}\over  
[\bar g_\ast]^{b_1+\beta^2\g^2 I_l\over 2 b_0}}  
$,
with $\bar g_\ast=  1+ {\bar z_\ast^d\over  z_0^d}$ is the geometrical factor
and all the expressions on the right are in terms of $\bar z_\ast$ only.
We may call the quantity ${b_1+\beta^2\g^2 I_l\over  b_0}$ as `geometric' index
of turning point.}
The last equation summarizes geometrically the whole effect of  IR bulk 
deformations (excitations), 
like having `black hole  in  geometry'  and  boosts
on the turning point value, perturbatively.

Having obtained the turning point expansion,
a similar expansion around  pure AdS  
can be made for the  area functional also.
After regularizing the area integral  \eqn{kl1kv}, in the UV limit
$(\epsilon \to 0 )$, 
we find the following expansion
\bea
 A&
\equiv & 2 \int_0^{z_\ast} {dz \over z^{d-1}}{K \over \sqrt{f} 
 \sqrt{K-K_\ast ({z\over z_\ast})^{ 2d-2}}} 
+ A_{UV}\br
& 
=& 2 \int_0^{z_\ast} {dz\over z^{d-1}}{1 \over  
 \sqrt{R}}[ 1+ {1+\beta^2\g^2\over 2}
{z^d\over z_0^d} + 
{\beta^2\gamma^2 \over 2}
 {(z_\ast^d-z^d)\over z_0^d}(-1+{1\over R}) +\cdots ] +A_{UV} 
\eea
where we have denoted diverging UV part as
$A_{UV}={2\over d-2}{1\over \epsilon^{d-2}}$.
The respective finite integrals can be evaluated at 
each order on the right hand side to give
\bea
 A&=&
{2\over z_\ast^{d-2}}[ a_0 + 
{z_\ast^{d}\over 2 z_0^d}( \g^2 a_1 + \beta^2\g^2 I_l)  
+\cdots] 
+A_{UV}\br
&&=
{2 a_0\over z_\ast^{d-2}}[ 1 + 
{z_\ast^{d}\over 2 z_0^d}( \g^2 {a_1\over a_0} 
+ {\beta^2\g^2\over a_0} I_l)  
+\cdots ] +A_{UV}
\eea
where new coefficients 
$a_0, a_1, ... $ are specific Beta-function integrals given in the appendix.
We should note that  $a_1>0$, but  using Beta function 
identities we shall have $a_0=-{b_0 \over d-2}$, so $a_0$ 
will be negative for all $d>2$.
Now substituting for $z_\ast$ from \eqn{dfg5} and only keeping terms up to 
first order we  find that
\bea\label{are1a}
A
&&= A_{UV}+
{2 a_0\over \bar z_\ast^{d-2}} \left(
1 
+ {\bar z_\ast^{d}\over  z_0^d}{d-2\over 2} ({b_1\over b_0}
 + \beta^2\g^2 {I_l\over  b_0})
  + 
{\bar z_\ast^{d}\over 2 z_0^d}( \g^2 {a_1\over a_0} 
+ {\beta^2\g^2}{ I_l\over a_0} ) \right) \br
&&= A_{UV}+ A_0
 \left( 1 
+ {\bar z_\ast^{d}\over  z_0^d}{d-2\over 2} {b_1\over b_0}+
{\bar z_\ast^{d}\over 2 z_0^d} \g^2 {a_1\over a_0} 
 \right) \br
&&\equiv A_{UV}+ A_0+A_1
  \eea
where in the second last line the
terms involving $I_l$  have got exactly cancelled! 
We have  defined 
\be
A_0={2 a_0\over \bar z_\ast^{d-2}} 
= 
- {(2b_0)^{d-1}\over (d-2) l^{d-2}} 
\ee
as the   
area contribution for pure $AdS_{d+1}$  with  turning point 
 $\bar z_\ast$ and strip width $l$. The term $A_1$
contains all the first order contributions to the area.
As a check,   for  
pure AdS ( $A_1=0$) we get the standard result \cite{RT}
\be \label{hj2}
A_{AdS}={1\over d-2}\left({2\over \epsilon^{d-2}} 
-{ 2^{d-1}b_0^{d-1}\over l^{d-2}}\right) .
\ee
which is a positive definite quantity.
From equation \eqn{are1a}  we  also find the 
net change in the area of extremal surface due to IR  deformations 
(black hole with boost).
It is given by
\bea\label{hj34}
\bigtriangleup A\equiv A-A_{AdS}&=& 
 {a_0\bar z_\ast^{2}\over  z_0^d} \left(
(d-2) {b_1\over b_0}+ \g^2 {a_1\over a_0} 
 \right) \br
&&= {a_1 l^{2}\over 4 b_0^2}
 \left(
{d-1\over d+1}+\beta^2\g^2 
 \right) {1\over  z_0^d} 
\eea
where in the second line we have used the
relation between two 
 ratios
 ${b_1\over b_0}=-{2\over (d+1)(d-2)} {a_1\over a_0} $. It is remarkable to
note that the remainder of the expression on the right hand 
side of eq.\eqn{hj34} 
is positive definite, which suggests that the net area of the extremal strip
 has effectively increased as compared to the
 pure AdS. The presence of $\beta$  dependent 
terms  precisely
contain the effect of  boost on the area of the extremal surface. 
In the absence of boost these terms will be absent and we shall get 
the result first obtained by \cite{JT}. It suggests that the
boosting of the bulk metric (which forms a type of charged excitations in the 
CFT$_d$)
 increases the strip area and hence increases the entanglement entropy
for the CFT subsystem. 
Following from \eqn{hj34} the change in 
entanglement entropy above  the pure AdS ground state, up to first order
is given by
\bea\label{hj4}
&& \bigtriangleup S  
= {L^{d-1} V_{d-2}\over 16 G_{d+1}} {a_1 l^2\over  b_0^2}
 \left(
{d-1\over d+1}+\beta^2\g^2 
 \right) {1\over  z_0^d} \ .
\eea
The equation \eqn{hj4}
is  an important expression for the remaining part of the analysis
in this section.

\subsection{Entanglement First Law}
It is  left now  to carefully partition the right hand side 
in terms of physical thermodynamic
observables of the CFT.
The physical quantities such as energy, 
 charge and pressure 
 can be obtained by expanding the bulk geometry \eqn{bst1} in
suitable Fefferman-Graham (asymptotic) coordinates near the AdS 
boundary \cite{fg}, given in the appendix. These for
the subsystem of CFT$_d$ (on a circle) are summarised here. The 
energy and charge for the strip subsystem are 
\bea\label{hj5}
&& \bigtriangleup {\cal E}={d L^{d-1} V_{d-2}l \over 16 \pi G_{d+1}} <t_{00}> 
=  {  r_yL^{d-1} V_{d-3}l \over 8 G_{d+1}}  
({d-1\over d} +\beta^2\gamma^2) {d\over
z_0^d}\br
&& \bigtriangleup {\cal N} \equiv r_y P_y
= { r_yL^{d-1} V_{d-2}l  \over 16 \pi G_{d+1}}  {\beta\gamma^2d\over z_0^d}
\eea
respectively. The pressure component along the $x_1$ direction of 
the compactified CFT is
\bea\label{hj5a}
&& \bigtriangleup {\cal P}= {2\pi r_y 
 L^{d-1}d  \over 16\pi G_{d+1}} <t_{11}> 
= {L^{d-1}r_y \over 8 G_{d+1} }  {1 \over z_0^d}
\eea
while $V_{d-3}\equiv l_2 l_3 \cdots l_{d-2}$, and 
$d$-dimensional Newton's constant  ${1\over G_d}={2\pi r_yL\over G_{d+1}}$. 
The ${\cal{N}}$ represents integral value of  
(momentum) charge. In the absence of boost it would be vanishing. We
 note down  nontrivial chemical potential 
in our solutions. It is given by
the value of gauge potential $\omega$ at the turning point,  
\be \label{chem1}
\mu ={1\over r_y\beta}(1-{1\over  K( z_\ast) })
\simeq {\beta\gamma^2\over r_y}{\bar z_\ast^d\over z_0^d} 
\ee  
Hence the contribution of `entanglement chemical potential'
 would remain negligible in first order of approximation
we are working in this section. 
(Note, the corresponding thermal  value of chemical potential 
is however  large   
  $\mu_{_{thermal}}={\beta\over r_y}$.) 

Our aim is to express the right hand side of \eqn{hj4} in terms of above 
physical observables. From \eqn{hj5},
a little guess tells us that 
  \bea
 \left(
{d-1\over d+1}+\beta^2\g^2 
 \right) {1\over  z_0^d}
\equiv  \bigg[({d-1\over d} +\beta^2\gamma^2)
-{d-1\over d+1} {1\over d}\bigg] {1\over z_0^d} 
\eea
Using \eqn{hj5} and \eqn{hj5a} we can now express eq.\eqn{hj4} as
\bea \label{alis1}
&& \bigtriangleup S_E  
= 
{1\over T_E} \left(
\bigtriangleup {\cal E}-
{d-1\over d+1} ~{\cal V}
\bigtriangleup {\cal P}
 \right) 
\eea
where $ {\cal V}
\equiv l [V_{d-3}] $ is the net volume of the subsystem. 
The equation \eqn{alis1} simply
 describes the  first law of entanglement thermodynamics, 
which is identical to the result in  
 \cite{alisha}. An alternative  first law
form was first proposed  by \cite{JT} for the isotropic AdS case.  
It leads to a difference in entanglement temperatures. If we set $\beta=0$
in \eqn{alis1}, it reduces to the known first law form obtained in 
\cite{alisha}.
Hence  we can conclude that the form of the first law
remains true  for  `boosted' AdS black-hole case as well, 
eventhough the   excitations in CFT
are much different  in  the boosted case. For example, there 
are  quantized charges present in these backgrounds.  
The entanglement temperature is given as
\be
T_E= { b_0^2\over a_1}{d\over \pi l}=
{ (B({d\over 2d-2},{1\over2}))^2 \over 2(d-1) 
B({1\over d-1},{1\over2})}{d\over \pi l}.
\ee
The temperature
 is inversely proportional to the width of strip. 
But this temperature is lower by a
factor  ${d\over d+1}$ as compared to the isotropic case
 in \cite{JT}.   
It is  evident that there is no explicit charge dependence 
in the first law equation \eqn{alis1}.
The reason for this is that the entanglement chemical potential given
in \eqn{chem1}
remains negligible $(\sim O(z_\ast^d /z_0^d))$ at the first order.
The contribution of chemical potential will however become important
in  next higher order calculation which we perform in the following section.
This contribution is expected to change the `first order' form of the first law \eqn{alis1}.

\section{Entanglement entropy  at second order}
Taking similar steps as in the previous section, we now calculate
the second order terms in the expansion of the area integral
schematically denoted as
\be
A\equiv A_{UV}+A_0+A_1 +A_2 
\ee
where $A_0$ and all first order terms contributing to
$A_1$ have been obtained in the previous section. Our aim is to find $A_2$.
First, we obtain the expansion of the turning point value, 
as done in \eqn{dfg5}, up to second order
\bea\label{dfg52}
 z_{\ast}&=&
  \bar z_\ast \left(  1 
+ {\bar z_\ast^{d}\over  z_0^d}( {b_1+ \beta^2\g^2 I_l \over 2 b_0})
+ {\bar z_\ast^{2d}\over  z_0^{2d}}\left({\bar b_2+ J_l \over  b_0}
-d( {b_1+ \beta^2\g^2 I_l \over 2 b_0})^2\right)
 \right)^{-1} 
\eea
where $\bar b_2\equiv {3\over 8}b_2$. The coefficients $b_2,~I_l,~J_l$  
are given  in the appendix. There is no need to simplify this 
expression further at this step.
A lengthy  calculation leads to  following
second order contribution 
\bea\label{ghy1}
&& A_2={2a_0\over \bar z_\ast^{d-2}}\bigg[
(d+3){(b_1 +\beta^2\gamma^2I_l)^2\over 8 a_0b_0}
-{(b_1 +\beta^2\gamma^2I_l)(\gamma^2 a_1+\beta^2\gamma^2I_l)\over 2a_0b_0}\br
&&~~~~~~~~~~~~
-{8\bar b_2- (3-\beta^2\gamma^2)\gamma^2 a_2\over 8a_0}
 +{1\over 2}\beta^4\gamma^4{I_a\over4 a_0}\bigg]
{\bar z_\ast^{2d}\over z_0^{2d}} \br
&& ~~~~ ={a_1\over \bar z_\ast^{d-2}}\bigg[
(d+3){(b_1 +\beta^2\gamma^2I_l)^2\over 4 a_1b_0}
-{(b_1 +\beta^2\gamma^2I_l)(\gamma^2 a_1+\beta^2\gamma^2I_l)\over a_1b_0}\br
&&~~~~~~~~~~~~
-{8\bar b_2- (3-\beta^2\gamma^2)\gamma^2 a_2\over 4a_1}
 +{1\over 4}\beta^4\gamma^4({I_a \over a_1})\bigg]
{\bar z_\ast^{2d}\over z_0^{2d}}
\eea
All parameters in the above expression are  known Beta functions
provided in the appendix. We need to further simplify the last equation. 
After some tedious simplifications  equation \eqn{ghy1}   can be rearranged as
\bea\label{ghy2}
&& A_2 ={a_1 \bar z_\ast^{2}}
\left(h_0+h_1 \beta^2\gamma^2+ h_2\beta^4\gamma^4\right)
{\bar z_\ast^{d}\over z_0^{2d}}
\eea
where coefficients are
\bea
&&
h_0={d-1\over d+1}(-{b_1\over2 b_0}+{3\over4}{d+1\over 2d+1}{a_2\over a_1}),\br
&& h_1=(-{b_1\over b_0}+{1\over2}{a_2\over a_1})\br
&&h_2={d+1\over d-1}(-{b_1\over 2 b_0}+{3\over4}{1\over d+1}{a_2\over a_1})\ .
\eea
Note the area integral $(A)$ is expanded around the AdS turning point. 
The  net change ($A_1+A_2$) in the area of the extremal strip 
 up to second order is given by
\bea\label{hj3c1}
\bigtriangleup A= 
{a_1 l^2\over 4 b_0^2}
 \left(
({d-1\over d+1}+\beta^2\g^2) 
  {1\over  z_0^d} 
+(h_0+h_1 \beta^2\gamma^2+h_2\beta^4\gamma^4){\bar z_\ast^{d}\over z_0^{2d}}
\right)\ .
\eea
At this point it is quite remarkable to notice that the equation \eqn{hj3c1} 
can also be written in an unique factorized form 
\bea\label{hj3c}
\bigtriangleup A
={a_1 l^{2} \over 4 b_0^2}\cdot Q \cdot \left( ({d-1\over d+1}+\beta^2\g^2) 
  {1\over  z_0^d} 
- {q a_2\over2 a_1}
 \beta^2\gamma^4
{\bar z_\ast^{d}\over z_0^{2d}}
\right)\ ,
\eea
where the  factor $Q$ (quotient) is given by\footnote{
It
is unique in the sense that after the  factorization
the remainder of the  expression in \eqn{hj3c}
(within large bracket)  precisely contains nontrivial
$\beta^2\gamma^4$ term, which
contributes to $\mu .\bigtriangleup {\cal N}$,
 alongwith usual  energy and pressure terms, as we would see next.
The `$Q$' factor is determined by simple quotienting procedure. Crucially 
there is no choice of $Q$ for which we can set $q=0$ in \eqn{hj3c}.
Any arbitrary $Q$ would take us back
to the situation where we started from, leaving us with little or no clue.
}
\bea
&&
Q= 1- 
\left((1+{d+1\over d-1}\beta^2\g^2){b_1\over2 b_0}
-(p+s{d+1\over d-1}\beta^2\g^2){a_2\over2 a_1}\right)
 {\bar z_\ast^{d}\over z_0^{d}}
\eea
with  unique set of parameters $p,~q,~s$ taking values as
\be\label{para5}
p={3\over 2}{d+1\over 2d+1},
 ~~~~s={2+8d-d^2\over 4(2d+1)}, ~~~~
q= {4+6d-d^2\over  4(2d+1)}\ .
\ee
The eq.\eqn{hj3c} is the complete expression representing the
net change in  area of the strip when calculated up to second order. 
From the  result \eqn{hj3c} we determine 
\bea\label{hj3cv}
&& \bigtriangleup S  
= {L^{d-1} V_{d-2}\over 16 G_N} 
 {a_1  l^{2} Q\over  b_0^2}  \left( ({d-1\over d+1}+\beta^2\g^2) 
  {1\over  z_0^d} 
- {qa_2\over2 a_1}
 \beta^2\gamma^4
{\bar z_\ast^{d}\over z_0^{2d}}
\right)
\eea
The eq.\eqn{hj3cv} provides  the complete expression representing the net
change in entanglement entropy up to second order in the expansion
around  pure AdS (ground state) value.

\subsection{Renormalization and  Entanglement First Law}
It is apparent from the expression \eqn{hj3cv} that
we  would have to define 
new `renormalized' quantities in order to have a first law like relation. 
We first introduce the renormalized
width of the strip as
\be
 l_R\equiv  Q^{1\over 2} l \ee
Since generally $0<Q<1$, the entanglement length decreases after second order
corrections. This would be true so long as 
we work within the pertubative regime.
Further we assume the principle \cite {JT} and propose that 
the new entanglement temperature is inversely 
proportional to the renormalized  width 
\bea\label{jn8}
T_E^\ast= 
{ db_0^2 \over \pi a_1  l_R }=
{T_E\over\sqrt{ Q}}
\eea
The $Q$  also introduces boost dependence 
in the entanglement temperature at the second order. Even if 
there is no boost ($\beta=0$), $Q$ would still be nontrivial.
With these definitions we
 redefine (renormalize) the `entanglement energy' 
and `entanglement charge'   for the subsystem 
(following from \eqn{hj5} and \eqn{hj5a})
\bea \label{hk5}
&&\bigtriangleup {\cal E}^\ast=
\sqrt{ Q}\bigtriangleup {\cal E},~~~~~~
\bigtriangleup {\cal N}^\ast=
\sqrt{Q}\bigtriangleup {\cal N}
\eea
and  redefine the entanglement volume  as
\be \label{hk5a}
 {\cal V}^\ast=
\sqrt{Q} {\cal V}=
\sqrt{Q} l V_{d-3}.
\ee
All above would simply  
happen provided we realize that the actual physical
size (width) of the subsystem  encountered by the excitations  
is $l_R$,
whereas the old  $l$ is just the `bare' (coordinate) size of  strip subsystem.
Since all  extensive thermodynamic quantities of the  subsystem
will depend on  strip width,
hence all expressions are renormalized by a single quantity
$Q$.
Finally we shall prefer to define `entanglement pressure' as 
\be
 {\cal P}^\ast
\equiv {d-1\over d+1} {\cal P}
\ee
while the `entanglement chemical potential' is 
\be\label{chem2}
\mu_\ast= {q \beta\gamma^2 \over r_y}{a_2\over2 a_1}
{\bar z_\ast^{d}\over z_0^{d}}\equiv 
 {4+6d-d^2\over 8(2d+1)} {a_2\over a_1}\mu
\ee
Note $\mu$ is the  turning point value given in \eqn{chem1}. 
From \eqn{hj3cv}  and using above expressions,
we find that the changes in entanglement  
entropy  up to second order can be expressed as 
\bea\label{alis2}
&& \bigtriangleup S_E^\ast  
= 
{1\over T_E^\ast} \left(
\bigtriangleup {\cal E}^\ast- \mu_\ast \bigtriangleup {\cal N}^\ast
- {\cal V}_\ast
\bigtriangleup {\cal P}^\ast
 \right) 
\eea
All thermodynamic quantities in the above result 
quantifying  excitations in the CFT subsystem are completely known. 

\noindent{\it Discussion:}\\
Let us discuss what we have achieved. We  introduced  `renormalized length'
for the subsystem
in order to retain the form of  first law of entanglement
thermodynamics. If we did not do so
we will have no hope of
having a first law like relation. Note that the bulk geometry is well defined
and the corresponding boundary energy-momentum tensor is also fixed.
Therefore, only option left for us is to look for correct  subsystem size. 
(The 
length $l\equiv 2 b_0 \bar z_\ast$ is good for `pure' AdS with turning
point value $\bar z_\ast$).
With the excitations in the CFT ($z_\ast$ being new turning point)
the relationship between old $l$ and $z_\ast$ is known at best 
perturbatively (order by order), through eq. \eqn{dfg52}.
But we can define new renormalized length $l_R$ at higher orders.
With the help of given expressions, the relationship between $l_R$ 
and $z_\ast$ can also be fixed, perturbatively, 
but is not needed in our results.
Thus, if $l$ is the size  at the first order
at the next order the
correct size becomes $l_R$.
Not only the  length, we have to correct the chemical potential
as well, remember the chemical potential is zero at the first order.
Other (extensive) thermodynamic quantities depend on the length,
so these also get  renormalized  once the
 size becomes $l_R$.
But, are these corrections quantum in nature? 
In  AdS/CFT we  deal with boundary CFT which is a strongly coupled
 quantum theory. Since we are expanding around pure
AdS (describing  CFT ground state), the small excitations of the CFT 
above the ground state will necassarily be "quantum" in nature.
These  excitations for small subsystem are controlled
by the smallness of the
 ratio  ${T_{th}\over T_E}$ or by the turning point to  horizon
ratio ${\bar z_\ast\over z_0}$. For example, in $d=4$ case, 
the dimensionless ratio
\be 
{\bar z_\ast^4\over z_0^4}
\propto g_{_{YM}}^4 {l^4} \epsilon_0 
\simeq {l^4\lambda^2\epsilon_0 
\over n^2} 
\ee
where  $\epsilon_0$ denotes  energy density of the excitations.
Thus the corrections to various entanglement 
quantities are quantum in nature and depend on perturbative 
Yang-Mills coupling constant $g_{_{YM}}$ (or the 't Hooft coupling 
 $\lambda\sim g_{_{YM}}^2n$).  

\noindent{\it Remarks for  $AdS_4$, $AdS_5$ and $AdS_7$:}\\
We note that the parameter $p,q,s$ in \eqn{para5}
are positive definite but  smaller than one
in  string/M-theory cases with $d=3,4$ and $d=6$.  
Also the two Beta-function ratios, ${b_1\over 2b_0}$ and 
${a_2\over 2a_1}$, are both positive definite and generally smaller than one.
The eq.\eqn{chem2} implies that entanglement chemical potential
is positive definite for these  conformal cases. Although the result in 
\eqn{chem2} is applicable for any   $d$ dimensions, but 
 for $d>6$, the parameter $q$ changes sign, hence
 the chemical potential $\mu_\ast$  will also change sign for $d>6$. 
This is a surprising result, but it
simply may be an indication of the fact that we are
going beyond the realm of applicability of string/M-theory.

\subsection{The $l$ dependent behaviour}
Let us make a few comments here. The boundary 
CFT is a $d$-dimensional theory having 
one of its  direction being compact. 
As there are  black holes in the bulk geometry 
it is a finite temperature theory. The thermal temperature is given by  
$T_{Th}= {d \over 4\pi z_0\gamma}$ which is fixed. Since the
 size of the subsytem is taken  small, so that the entanglement effects 
can be studied perturbatively, it leads to a hierarchy of scales
\be
{\bar z_\ast\over z_0}\ll 1,~~~~
{l\over z_0}\ll 2b_0,~~~~{T_{Th}\over T_E}\ll {a_1\over2 b_0\gamma}
\ee
while  we keep $\beta\gamma \sim 1$.
The  renormalized temperature \eqn{jn8}
up to second order can  be written as
\bea\label{kl8}
T_E^\ast\simeq 
{ 1\over \pi a_1  l}{d b_0^2\over \sqrt{1 -\alpha_0 
({2\pi\gamma l T_{Th} \over db_0})^d}}
\eea
where $\alpha_0 \equiv
\left((1+{d+1\over d-1}\beta^2\g^2){b_1\over2 b_0}
-(p+s{d+1\over d-1}\beta^2\g^2){a_2\over2 a_1}\right)$ is always
positive definite. 
This expression remains valid so long as ${2\pi\gamma l T_{Th} \over b_0d}<1$ 
is maintained.
The eq. \eqn{kl8}
 implies that the entanglement temperature  has sizable
 corrections for large $l$ from higher order at a given  thermal temperature
$T_{Th}$. It also tells us how the entanglement temperature will
flow towards  $T_{Th}$ as $l$ increases. From \eqn{kl8}, while
keeping the strip size $l$ fixed, we can also study 
the flow of entanglement temperature
with respect to change in (black hole) thermal temperature
\bea\label{Kkl8}
T_E^{(2)} \simeq 
{T_E^{(1)} 
\over \sqrt{1 -\alpha_0 
({2\pi\gamma l \over db_0})^d 
(T_{Th}^{(2) d}- T_{Th}^{(1)d})}}
\eea
where $T_{Th}^{(2)}$ and $ T_{Th}^{(1)}$ are two different
 black hole temperatures. The equation \eqn{Kkl8} implies
that the entanglement temperature will be higher for
the bigger size black hole ($T_{Th}^{(2)}> T_{Th}^{(1)}$).
The `$T_E$ Vs $l$' graphs have been plotted in the figure \eqn{figure4}
for different $T_{Th}$ values. 
\begin{figure}[h]
\centerline{\epsfxsize=3.5in
\epsffile{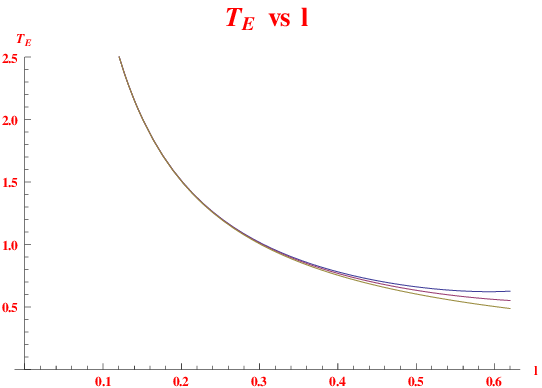} }
\caption{\label{figure4} 
\it Plots of `$T_E$ Vs $l$' for different black 
hole temperatures (starting from top curve)
$T_{Th}=.28, .25, \& .10$ with fixed $\alpha_0=.97$ and 
$(\beta\gamma)^2=.5$ for  $AdS_5$. The graphs split at large $l$ showing the
effect of corrections. These demonstrate  that $T_E$ is
higher for  higher black hole temperature.  
}
\end{figure}
 
The entanglement energy  of  subsytem 
gets corrected as
\bea
&&
\bigtriangleup {\cal E}^\ast=  
\sqrt{1 -{\alpha_0} ({2\pi\gamma l T_{Th}\over db_0})^d} 
\bigtriangleup {\cal E} 
\br &&~~~~=  
l\sqrt{1 -{\alpha_0} ({2\pi\gamma l T_{Th}\over db_0})^d} 
{ L^{d-1} V_{d-3}r_yd \over 8 G_{d+1}} <t_{00}> 
\eea
From \eqn{chem2} the chemical potential up to the second order  
may be written
\bea
&& \mu_\ast= {qa_2\over 2r_y a_1} 
{\beta\gamma^2 ({2\pi
\gamma l T_{Th}\over db_0})^d} 
+ {\rm~higher~orders} \br
&& ~~~~\simeq {qa_2\over 2r_y\beta a_1} (1-
{1\over{1 +\beta^2\gamma^2 ({2\pi
\gamma l T_{Th}\over db_0})^d}}) 
\eea
where the second line merely reflects the fact that any
subleading term is a higher order term which  can be ignored
at the second order.
This will lead to following $l$ dependence in the charge 
\bea && \bigtriangleup {\cal N}^\ast 
=l\sqrt{1 -{\alpha_0 } ({2\pi
\gamma l T_{Th}\over db_0})^d}~ { L^{d-1} V_{d-3}r_y^2d  
\over 8 G_{d+1}}  {\beta\gamma^2\over z_0^d}
\eea
The large $l$ behaviour may be predicted from here
 up to some value $l=l_c$, such that  $l_c  
< {d b_0\over 2\pi\gamma  T_{Th}}$. 
We cannot stretch these results beyond this bound as this would lead to
  to the break down of perturbative regime. 
In  large $l$ limit
we   expect to see the  behaviour
\be
T_E^\ast\to T_{Th},  ~~~~\bigtriangleup {\cal E}^\ast \to \bigtriangleup 
{\cal E}_{Th}.
\ee

\section{Summary}
We conclude that 
the first law of entanglement thermodynamics for `boosted' $AdS_{d+1}$
 having black hole in the IR region is given by
\bea\label{alis2a}
&& 
\bigtriangleup {\cal E}^\ast
= T_E^\ast \bigtriangleup S_E^\ast  
 + \mu_\ast \bigtriangleup {\cal N}^\ast
+ {\cal V}_\ast
\bigtriangleup {\cal P}^\ast \nonumber
\eea
Our result emphasizes the fact that the form 
of the first law  changes under  higher order corrections
to the entanglement entropy. It is apparent when the entanglement
law \eqn{alis1} at the first order
 is compared with the second order result in \eqn{alis2}. 
We find that even in the absence of boosts the
 renormalization of the thermodynamic quantities 
like entropy, energy, subsystem size (all extensive quantities) 
and entanglement temperature (intensive quantity)  
becomes essential at the second order. The chemical potential 
which is negligible at the first order  becomes relevant
 at next order. 
We expect no further changes in the form 
of the first law for the AdS background  \eqn{bst1}, so
 the first law form \eqn{alis2} will remain unchanged
at higher orders provided we renormalize/redefine the
thermodynamic quantities appropriately. 
Also, we have determined that the entanglement temperature 
of the subsystem will be higher for
a bigger size black hole. 
Finally, as we have 
 studied (IR) excitations in  AdS spacetime, and 
since AdS background is an universal solution
of (gauged) supergravities with negative cosmological constant, 
we expect these results  will be holding true quite generally.

%\newpage
%\vskip.5cm
%\noindent{\it Acknowledgments:} 

\vskip.5cm
   
\appendix{

\section{ Conventions:}

The physical observables such as enegry, momentum and pressure 
 can be obtained by expanding 
the bulk AdS geometry \eqn{bst1} in
suitable Feffermann-Graham asymptotic coordinates \cite{fg}
\bea\label{yt1}
&&ds^2={L^2\over u^2}
\left( {du^2}+G^{4\over d}[{-f dt^2\over K}+K (dy-\omega)^2
+dx_1^2+\cdots+dx_{d-2}^2]\right) \br
&& G= 1+{u^d\over u_0^d},~~~f\simeq (1-{4u^d\over u_0^d}),
~~~K\simeq 1+4\beta^2\gamma^2{u^d\over u_0^d} 
\eea
In $u$ coordinate the boundary is at $u=0$, and  $u_0^d = 4 z_0^d$. The
Kaluza-Klein gauge form is 
\be
\omega=\beta^{-1}(1-{1\over K}) dt.
\ee
In these asymptotic coordinates, 
the coefficients of  ${u^d}$ terms in the  metric expansion give rise
to the energy-momentum tensor of the boundary CFT. From \eqn{yt1}
these coefficients of the metric are
\bea
&&<t_{00}>=
({d-1\over d} +\beta^2\gamma^2){4\over u_0^d}, ~~~~<t_{0y}>=
\beta\gamma^2{4\over u_0^d}\br
 &&<t_{11}>=
{1\over d}{4\over u_0^d}=<t_{22}>= ~\cdots
\eea
The boundary
energy-momentum tensor, $<T_{ab}>={dL^{d-1}\over 16\pi G} <t_{ab}>$, is traceless as we have  conformal  theory.
The energy of excitations and the momentum  for
the boosted CFT$_d$ will be
\bea
&&  E= { d L^{d-1} v_{d-1}\over 16 \pi G_{d+1}} <t_{00}> 
= { d L^{d-1} v_{d-2} r_y\over 8  G_{d+1}}  
({d-1\over d} +\beta^2\gamma^2)z_0^{-d}\br
&&  P_y
= { d L^{d-1} v_{d-2}r_y \over 8 G_{d+1}}  \beta\gamma^2z_0^{-d}
\eea
where volume $v_{d-2}=l_1 l_2 \cdots l_{d-2}$, and
 we have  compactified   $ y$  
on a circle of radius $r_y$. Note the momentum (charge) $P_y={N\over r_y}$ 
is quantized and $N$ would have integral values. 
 In the absence of boost the charge would be vanishing. 
We note down the nontrivial chemical potential 
 which is defined by
the value of gauge potential at the horizon  
\be
\mu_{Th} 
={\beta\over r_y}
\ee  
Corresponding thermal entropy and temperature can be obtained 
from \eqn{bst1}. These are given by
\bea 
&&
S_{Th}\equiv {[Area]_{horizon}\over 4G_{d+1}} 
= {\pi L^{d-1} v_{d-2}r_y\over 2 G_{d+1} } {\gamma \over z_0^{d-1}} \br
&& T_{Th}= {d\over 4\pi z_0\gamma}
\eea
These thermal quantities 
satisfy the following first law of black hole mechanics
\be
\delta E_{Th}= T_{Th}\delta S_{Th} + \mu_{_{Th}}\delta N  \ . 
\ee
But if we allow small volume changes, say $\delta v=(\delta l_1)l_2l_3\cdots l_{d-2}$,
the black hole thermodynamic law would be
\be
\delta E_{Th}= T_{Th}\delta S_{Th} + \mu_{_{Th}}\delta N -{\cal P}_1\delta v \ . 
\ee
where pressure component is $  {\cal P}_1
= {  L^{d-1} r_y \over 8 G_{d+1}}  z_0^{-d}$.

\section{Some Beta function Identities}
Some useful Beta function integrals we have used are given here
\bea
&& b_0=\int_0^{1} d\xi \xi^{d-1}{1 \over   \sqrt{R}}
={1\over 2(d-1)}B({d\over2d-2},{1\over2})\br
&& b_1=\int_0^{1} d\xi \xi^{2d-1}{1 \over   \sqrt{R}}
={1\over 2(d-1)}B({d\over d-1},{1\over2})\br
&& b_2=\int_0^{1} d\xi \xi^{3d-1}{1 \over   \sqrt{R}}
={1\over 2(d-1)}B({3d\over 2d-2},{1\over2})\br
&& I_l=\int_0^{1} d\xi \xi^{d-1}(1-\xi^d){1 \over   {R}^{3\over2}}=
{d+1\over d-1} b_1
-{1\over d-1} b_0 \br
&& J_l=\int_0^{1} d\xi \xi^{d-1} \left( 
{\beta^2\gamma^2\over 4} \xi^d
-{\beta^4\gamma^4\over 8} (1+3\xi^d)\right)
{(1-\xi^d) \over   {R}^{3\over2}}
\eea
where $B(m,n)={\Gamma(m)\Gamma(n)\over\Gamma(m+n)}$ are the Beta-functions.
Further integrals are
\bea
&& a_0=\int_0^{1} d\xi \xi^{-d+1}{1 \over   \sqrt{R}}
={1\over 2(d-1)}B({1-d/2\over d-1},{1\over2})\br
&& a_1=\int_0^{1} d\xi \xi^{-d+1}{\xi^d \over   \sqrt{R}}
={1\over 2(d-1)}B({1\over d-1},{1\over2})\br
&& a_2=\int_0^{1} d\xi \xi^{-d+1}{\xi^{2d} \over   \sqrt{R}}
={1\over 2(d-1)}B({1+d/2\over d-1},{1\over2})\br
&& I_a=\int_0^{1} d\xi \xi^{d-1}(1-\xi^{2d}){1 \over   {R}^{3/2}}
={2d+1\over d-1} b_2
-{1\over d-1} b_0
\eea
Some 
identities we  have used are 
\bea
b_0=(2-d) a_0, ~~~~
b_1={2\over d+1} a_1, ~~~~
b_2={2+d\over 2d+1} a_2 \ .
\eea

}
\vskip.5cm
%\newpage

\end{document}